\documentclass[british,aps,prl,superscriptaddress]{revtex4}
\usepackage[T1]{fontenc}
\usepackage[latin9]{inputenc}
\usepackage{verbatim}

\usepackage{array}
\usepackage{float}
\usepackage{textcomp}
\usepackage{amsmath}
\usepackage{amssymb}
\usepackage{graphicx}
\usepackage{hyperref}

\usepackage{color,soul}

\makeatletter

\providecommand{\tabularnewline}{\\}
\@ifundefined{textcolor}{}
{%
 \definecolor{BLACK}{gray}{0}
 \definecolor{WHITE}{gray}{1}
 \definecolor{RED}{rgb}{1,0,0}
 \definecolor{GREEN}{rgb}{0,1,0}
 \definecolor{BLUE}{rgb}{0,0,1}
 \definecolor{CYAN}{cmyk}{1,0,0,0}
 \definecolor{MAGENTA}{cmyk}{0,1,0,0}
 \definecolor{YELLOW}{cmyk}{0,0,1,0}
}

\usepackage{graphicx}  % needed for figures
\usepackage{dcolumn}   % needed for some tables
\usepackage{bm}        % for math

\makeatother

\usepackage{babel}
\begin{document}

\title{Effect of Loss on Multiplexed Single-Photon Sources}

\author{Damien Bonneau\footnotemark\footnotetext{Authors DB and GJM contributed equally to this work.}}

\affiliation{Centre for Quantum Photonics, H. H. Wills Physics Laboratory \& Department of Electrical and Electronic Engineering, University of Bristol, Merchant Venturers Building, Woodland Road, Bristol, BS8 1UB, UK}

\author{Gabriel J. Mendoza\footnotemark[\value{footnote}]}

\affiliation{Centre for Quantum Photonics, H. H. Wills Physics Laboratory \& Department of Electrical and Electronic Engineering, University of Bristol, Merchant Venturers Building, Woodland Road, Bristol, BS8 1UB, UK}

\author{Jeremy L. O'Brien}

\affiliation{Centre for Quantum Photonics, H. H. Wills Physics Laboratory \& Department of Electrical and Electronic Engineering, University of Bristol, Merchant Venturers Building, Woodland Road, Bristol, BS8 1UB, UK}

\author{Mark G. Thompson\footnotemark\footnotetext{mark.thompson@bristol.ac.uk}}

\affiliation{Centre for Quantum Photonics, H. H. Wills Physics Laboratory \& Department of Electrical and Electronic Engineering, University of Bristol, Merchant Venturers Building, Woodland Road, Bristol, BS8 1UB, UK}

\begin{abstract}
An on-demand single-photon source is a key requirement for scaling
many optical quantum technologies. A promising approach to realize
an  on-demand single-photon source is to multiplex an array of heralded
single-photon sources using an active optical switching network. However,
the performance of multiplexed sources is degraded by photon loss
in the optical components and the non-unit detection efficiency of
the heralding detectors. We provide a theoretical description of a
general multiplexed single-photon source with lossy components and
derive expressions for the output probabilities of single-photon emission
and multi-photon contamination. We apply these expressions to three
specific multiplexing source architectures and consider their tradeoffs
in design and performance. To assess the effect of lossy components
on near- and long-term experimental goals, we simulate the multiplexed
sources when used for many-photon state generation under various amounts
of component loss. We find that with a multiplexed source composed
of switches with $\sim0.2-0.4$ dB loss and high efficiency number-resolving
detectors, a single-photon source capable of efficiently producing
20-40 photon states with low multi-photon contamination is possible,
offering the possibility of unlocking new classes of experiments and
technologies.
\end{abstract}
\maketitle

\section*{Introduction}

An on-demand single-photon source is a key requirement for many optical
quantum technologies, including quantum key distribution schemes using
quantum repeaters \cite{Sangouard-key_dist_2012}, quantum metrology
using photon-number states \cite{Cable-PRL-2007}\cite{Kishore-metrology-2005},
analog quantum simulators \cite{Aspuru-Guzik:2012aa}, and the boson
sampling machine \cite{Motes-boson-2013}. The ultimate optical quantum
technology may be the quantum computer \cite{Knill:2001aa}, capable
of efficient integer factorization and digital quantum simulation
\cite{Shor:1997:PAP:264393.264406}\cite{Lloyd23081996}, which relies
critically on the development of a high-performance, on-demand photon
source in order to efficiently generate large quantum resource states
\cite{gr-njp-2004}\cite{Varnava-PRL-2008}.

\begin{figure}
\begin{centering}
\includegraphics[scale=0.2]{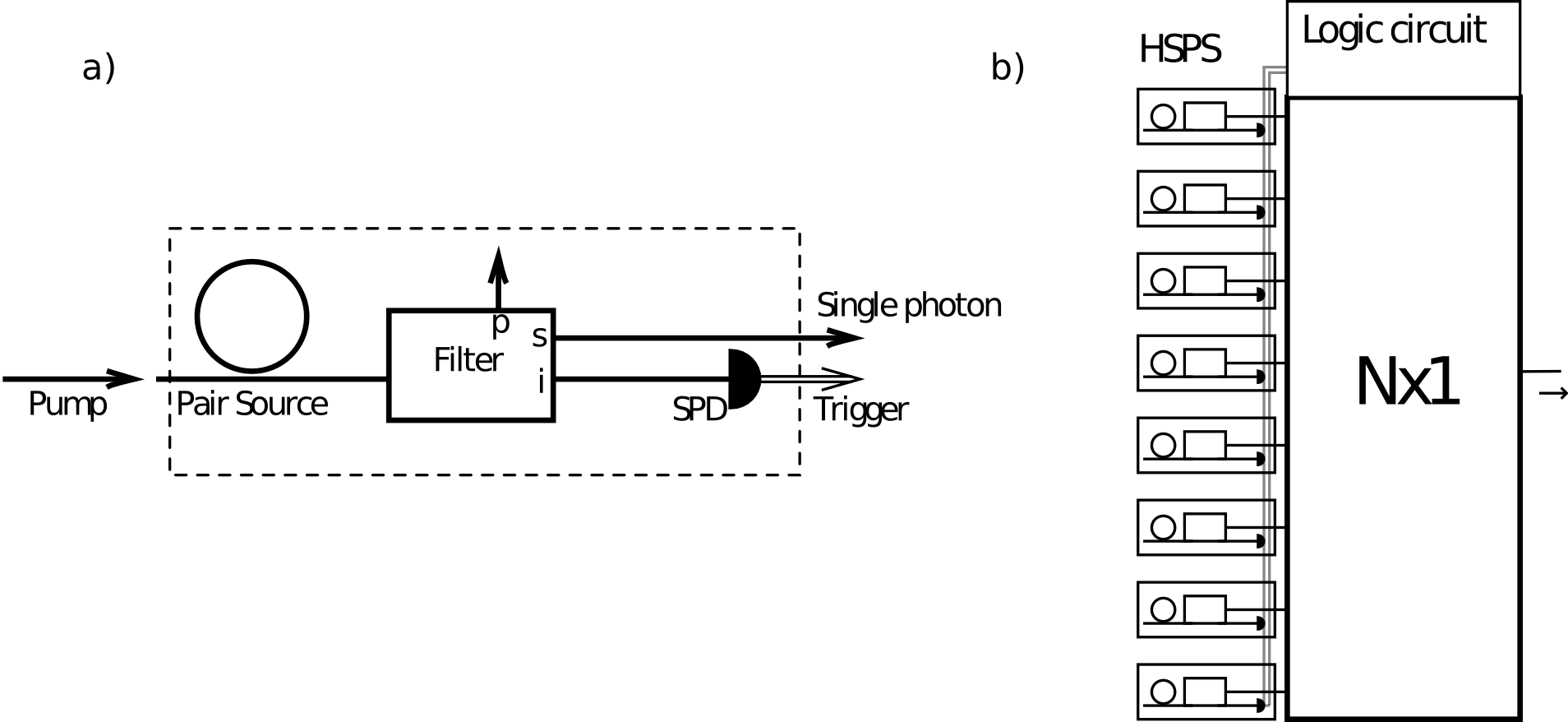}
\par\end{centering}

\protect\caption{a) Heralded single-photon source (HSPS). A laser pumps a parametric
pair source which spontaneously emits a signal and an idler photon.
A filter separates the pump, signal, and idler. The idler photon is
detected using a single-photon detector (SPD), heralding the presence of
the signal photon. b) A general multiplexed (MUX) single-photon source.
$N$ HSPSs are pumped simultaneously; the idler photons are detected
while the signal photons are stored in a long delay line. A classical
logic unit determines the configuration for the $N\times1$ switching
network based on the detection signals, routing a successfully generated
single photon to the output.}\label{fig:HSPSandMUX}
\end{figure}

While research efforts continue in developing on-demand single-photon
sources using atomic systems, achieving high spectral purity and high
collection efficiency simultaneously remains a challenge \cite{Eisaman-review-single-2011}.
An alternative approach commonly used in quantum optics experiments
is heralded single-photon sources (HSPSs) such as those based on spontaneous
parametric down-conversion (SPDC) or spontaneous four-wave mixing
(SFWM): parametric processes which use a pump laser in a nonlinear
material to spontaneously generate photon pairs (Fig. \ref{fig:HSPSandMUX}a). Detecting
one of the photons using a single-photon detector will herald the
presence of the paired photon to be input into the quantum circuit.
Parametric sources are capable of producing single photons with high
spectral purity \cite{Eisaman-review-single-2011}, benefit from a
well-defined wave vector leading to high collection efficiency \cite{cu-ol-38-1609},
and have no mode-mismatch when integrated on the same monolithic substrate
as the subsequent circuit. However, parametric sources are inherently
inefficient, due to the thermal nature of the output statistics and
the requirement to keep the multi-photon emissions low, resulting
in a maximum single-photon output probability of 25\% \cite{silberhorn_multiplexedPDC_2012}.
For applications requiring many single photons, the probability of
successfully generating $N$ single photons simultaneously with $N$
HSPSs decreases as $N$ becomes large, severely limiting the size
of practical circuits. 

An approach to overcome all of the scaling problems with HSPSs is
the \textit{multiplexed} (MUX) single-photon source \cite{Migdall-pra-66-053805},
which uses an array of HSPSs, delay lines, electronics for classical
logic operations, and an active optical switching network to approximate
a true on-demand source (Fig. \ref{fig:HSPSandMUX}b). Using an array of HSPSs as a
collective unit means that the probability that at least one of the
HSPSs emits a single photon is high, while a switching network driven
by the heralding signals is used to route the generated photon into
a specific spatial-temporal output mode. This results in the near-deterministic
generation of single photons, allowing for much larger quantum circuits
than could be feasibly built with HSPSs without active multiplexing.
MUX sources inherit the same benefits as parametric HSPSs, including
their mature theoretical and experimental investigation, and are especially
appealing due the prospects of a fully integrated device with existing
fabrication processes.

Several theoretical schemes have been previously investigated \cite{br-oe-19-22698,gl-apl-103-031115,Jeffrey2004,kw-103-163602,Pittman-pra-66-042303,Shapiro-multi-optlet-2007,Mazzarella-pra-2013,Mower-PRA-multiplexed-2011,Schmiegelow:2014aa}
and experimental work using bulk \cite{ZeiA-pra-multi-2011} and integrated
\cite{Collin-Nat-2013}\cite{LPOR:LPOR201400027} components has been demonstrated. Previous
work has highlighted the fundamental constraints for creating pure
states using parametric processes assuming ideal MUX components \cite{silberhorn_multiplexedPDC_2012}.
However, in real physical settings, non-ideal components will limit
the efficiency and output fidelity of MUX sources. The most significant
sources of error in multiplexed sources are likely to be photon loss
in the optical components and the non-unit detection efficiency of
the heralding detectors (which can also be viewed as photon loss).
Assessing the suitability of a MUX sources using lossy components
is an important consideration for many quantum photonic technologies,
in particular the large-scale quantum computer, as fault-tolerance
thresholds will place demands on the required source efficiency and
fidelity \cite{Varnava-PRL-2008}\cite{Hayes-Parity-PRA-2010}. 

The aim of this work is to assess the impact of photon loss on the
performance of multiplexed single-photon sources. We start by detailing
the HSPS, which is a key building block of all the MUX sources, and
include the effect of photon loss and non-unit efficiency number-resolving
and non-number resolving detectors. We extend this description to
general multiplexed single-photon sources with lossy components and
derive expressions for single-photon and multi-photon emission probabilities.
We then apply these expressions to three specific multiplexing source
architectures and consider their tradeoffs in design and performance.
To assess the effect of lossy components on near- and long-term experimental
goals, we simulate the multiplexed sources  when used for many-photon
state generation under various amounts of component loss. We conclude
by discussing the prospects of using the different MUX architectures
with realistic components for near- and long-term experimental goals.

\section{The heralded single photon source\label{sec:The-heralded-single}}

\subsection{Theory and figures of merit}

All the MUX sources we consider are composed of a core component called
the heralded single photon source (HSPS) (Fig. \ref{fig:HSPSandMUX}a). A HSPS is a non-deterministic
source with a logic output set to 1 when it emits a photon and set
to 0 otherwise. We consider a HSPS composed of 1) a photon pair generation
stage in which photons are produced at non-degenerate wavelengths
$\lambda_{s}$ (signal) and $\lambda_{i}$ (idler), 2) a filter which
removes the pump and separates the signal from the idler into two
different paths, and 3) a detector on the idler arm which heralds
the emission of the signal photon. For now, we assume the photon pairs
are generated as a biphoton two-mode squeezed vacuum state such that
their joint spectrum is disentangled \cite{Grice2001}. Furthermore,
we assume that all signal photons produced from different HSPSs are
perfectly indistinguishable in all degrees of freedom except for the
spatial mode in which they are generated.

The source is characterised by a squeezing parameter $\xi$, determined
by the pump intensity and strength of the nonlinearity, a trigger
probability $p_{trig}$, the probability for the heralded state to
be a single photon $p_{single}$, also called the state fidelity or
heralding efficiency, and the probability for the heralded state to
be contaminated with multiple photons, $p_{multi}$. Quantifying multi-photon
contamination separately from vacuum emissions is especially important; vacuum emissions can be treated as effective loss in a linear
optical quantum circuit while multi-photon events can result in other
types of errors, and linear optical quantum circuits have been shown
to have a much higher tolerance to loss than to other errors \cite{Varnava-PRL-2008}.

We aim at deriving expressions for $p_{trig}$, $p_{single}$ , and
$p_{multi}$ for the case in which the heralding detector is number-resolving
and the case in which the detector is non-number resolving (also called
a threshold detector). The state produced by the pair source, assuming
spectral disentanglement, is of the form \cite{Gerry-Knight-QO}:

\begin{equation}
\left|\psi\right\rangle =\sqrt{1-\left|\xi\right|^{2}}\left(\left|0\right\rangle _{i}\left|0\right\rangle _{s}+\xi\left|1\right\rangle _{i}\left|1\right\rangle _{s}+\sum_{n=2}^{\infty}\xi^{n}\left|n\right\rangle _{i}\left|n\right\rangle _{s}\right),\label{eq:SPDC state}
\end{equation}
 where $i$ and $s$ are the idler and signal modes. In practice,
sources and filters have losses, and the heralding detector does not
have a unit efficiency detection. We call $\eta_{i}$ the global collection
efficiency on the idler arm accounting for all these effects, and
$\eta_{s}$ is the overall transmission on the signal arm accounting
for losses in the sources and filters. Each lossy component is modelled
as an ideal component preceded by an ideal beamsplitter with a non-unit
transmission probability. The full state, after accounting for losses
and tracing over loss modes, can therefore be written:

\begin{equation}
\noindent\hat{\rho}=\left(1-\left|\xi\right|^{2}\right)\left(\sum_{n=0}^{\infty}\left|\xi\right|^{2n}\sum_{p=0}^{n}\sum_{k=0}^{n}C_{n}^{p}\eta_{i}^{p}\left(1-\eta_{i}\right)^{n-p}C_{n}^{k}\eta_{s}^{k}\left(1-\eta_{s}\right)^{n-k}\hat{\rho}_{p,k}\right),\label{eq:pair genration reduced state}
\end{equation}
where $\hat{\rho}_{p,k}=\left|p\right\rangle _{i}\left|k\right\rangle _{s}\left\langle p\right|_{i}\left\langle k\right|_{s}$
and $C_{n}^{k}$ is the binomial coefficient.

This expression will be the starting point to derive $p_{trig\, D}$,
$p_{single\, D}$, and $p_{multi\, D}$ for the two detection schemes.
In this paper we use the subscript $D$ on expressions to indicate
the type of detector considered: NRD for number-resolving detector
and TD for threshold detector. Detailed derivations can be found in
Appendix A.

\subsection{Heralded single-photon source parameters}

\subsubsection*{Threshold Detector}

Starting with the reduced state (Eq. \ref{eq:pair genration reduced state})
and tracing out the signal mode, the probability for the detector
on the idler arm to trigger is given by summing all the contribution
of the states having at least one photon and is given by:

\begin{equation}
p_{trig\, TD}=\frac{\left|\xi\right|^{2}\eta_{i}}{1-\left|\xi\right|^{2}\left(1-\eta_{i}\right)}.
\end{equation}

The heralded state in the signal arm is expressed, renormalizing by
dividing by $p_{trig\, TD}$, as: 

\begin{equation}
\noindent\hat{\rho}_{heralded\, TD}=\frac{\left(1-\left|\xi\right|^{2}\right)}{p_{trig\, TD}}\left(\sum_{n=1}^{\infty}\left|\xi\right|^{2n}\sum_{p=1}^{n}\sum_{k=0}^{n}C_{n}^{p}\eta_{i}^{p}\left(1-\eta_{i}\right)^{n-p}C_{n}^{k}\eta_{s}^{k}\left(1-\eta_{s}\right)^{n-k}\left|k\right\rangle _{s}\left\langle k\right|_{s}\right).
\end{equation}

We can compute the probability that the heralded state is a single-photon
using $p_{single\, TD}=\left\langle 1\right|_{s}\hat{\rho}_{heralded\, TD}\left|1\right\rangle _{s}$:

\begin{equation}
\noindent p_{single\, TD}=\left(1-\left|\xi\right|^{2}\right)\eta_{s}\frac{\left[1-\left(\left|\xi\right|^{2}\left(1-\eta_{s}\right)\right)^{2}\left(1-\eta_{i}\right)\right]\left[1-\left|\xi\right|^{2}\left(1-\eta_{i}\right)\right]}{\left[1-\left|\xi\right|^{2}\left(1-\eta_{s}\right)\right]^{2}\left[1-\left|\xi\right|^{2}\left(1-\eta_{s}\right)\left(1-\eta_{i}\right)\right]^{2}}.\label{eq:p_single_TD}
\end{equation}

The probability that the heralded state contains multi-photon contamination
is given by

$p_{multi\, TD}=\sum_{m=2}^{\infty}\left\langle m\right|_{s}\hat{\rho}_{heralded\, TD}\left|m\right\rangle _{s}$:

\begin{equation}
p_{multi\, TD}=\frac{Z_{TD}}{p_{trig\, TD}}-p_{single\, TD},\label{eq:p_multi_TD}
\end{equation}

with

\begin{equation}
Z_{TD}=\left(1-\left|\xi\right|^{2}\right)\left|\xi\right|^{2}\left(\frac{1}{1-\left|\xi\right|^{2}}+\frac{\left(1-\eta_{s}\right)\left(1-\eta_{i}\right)}{1-\left|\xi\right|^{2}\left(1-\eta_{s}\right)\left(1-\eta_{i}\right)}-\frac{\left(1-\eta_{i}\right)}{1-\left|\xi\right|^{2}\left(1-\eta_{i}\right)}-\frac{\left(1-\eta_{s}\right)}{1-\left|\xi\right|^{2}\left(1-\eta_{s}\right)}\right).
\end{equation}
Note that we can also similarly derive an expression for the probability
that the heralded state is any Fock number state using $\left\langle n\right|_{s}\hat{\rho}_{heralded\, TD}\left|n\right\rangle _{s}$.

\subsubsection*{Number-resolving detector}

We now consider the case in which the detector is number resolving.
In practice, we only need to discriminate the single-photon state
from the vacuum and from states having two or more photons. Therefore,
the detector does not need to be able to distinguish between two and
$N>2$ photon states. Starting again from Eq. \ref{eq:pair genration reduced state},
tracing out the signal mode and calculating the probability to get
only one photon gives:

\begin{equation}
p_{trig\, NRD}=\frac{\left(1-\left|\xi\right|^{2}\right)\left|\xi\right|^{2}\eta_{i}}{\left(1-\left(1-\eta_{i}\right)\left|\xi\right|^{2}\right)^{2}}.
\end{equation}
The corresponding heralded state is:

\begin{equation}
\hat{\rho}_{heralded\, NRD}=\frac{\left(1-\left|\xi\right|^{2}\right)}{p_{trig\, NRD}}\eta_{i}\left(\sum_{n=1}^{\infty}\left|\xi\right|^{2n}n\left(1-\eta_{i}\right)^{n-1}\sum_{k=0}^{n}C_{n}^{k}\eta_{s}^{k}\left(1-\eta_{s}\right)^{n-k}\left|k\right\rangle _{s}\left\langle k\right|_{s}\right).
\end{equation}
As in the previous case, we can compute the probability the heralded
state contains one photon $\noindent p_{single\, NRD}=\left\langle 1\right|_{s}\hat{\rho}_{heralded\, NRD}\left|1\right\rangle _{s}$:

\begin{equation}
p_{single\, NRD}=\left(1-\left(1-\eta_{i}\right)\left|\xi\right|^{2}\right)^{2}\eta_{s}\left(\frac{\left(1+\left(1-\eta_{i}\right)\left(1-\eta_{s}\right)\left|\xi\right|^{2}\right)}{\left(1-\left(1-\eta_{i}\right)\left(1-\eta_{s}\right)\left|\xi\right|^{2}\right)^{3}}\right).\label{eq:p_single_NRD}
\end{equation}
Not surprisingly, contrary to the previous case, number-resolving
detectors enable unit fidelity in the absence of losses ($p_{single\, NRD}=1$
when $\eta_{i}=\eta_{s}=1$). The probability for the heralded state
to contain multi-photon contamination is given by:

\begin{equation}
p_{multi\, NRD}=\eta_{s}\frac{1-\left(1-\eta_{s}\right)\left(\left|\xi\right|^{2}\left(1-\eta_{i}\right)\right)^{2}}{\left(1-\left|\xi\right|^{2}\left(1-\eta_{i}\right)\left(1-\eta_{s}\right)\right)^{2}}-p_{single\, NRD}.\label{eq:p_multi_NRD}
\end{equation}
We can again similarly derive an expression for the probability that
the heralded state is any Fock number state using $\left\langle n\right|_{s}\hat{\rho}_{heralded\, NRD}\left|n\right\rangle _{s}$.

\subsection{Discussion }

\begin{figure}
\begin{centering}
\includegraphics[scale=0.85]{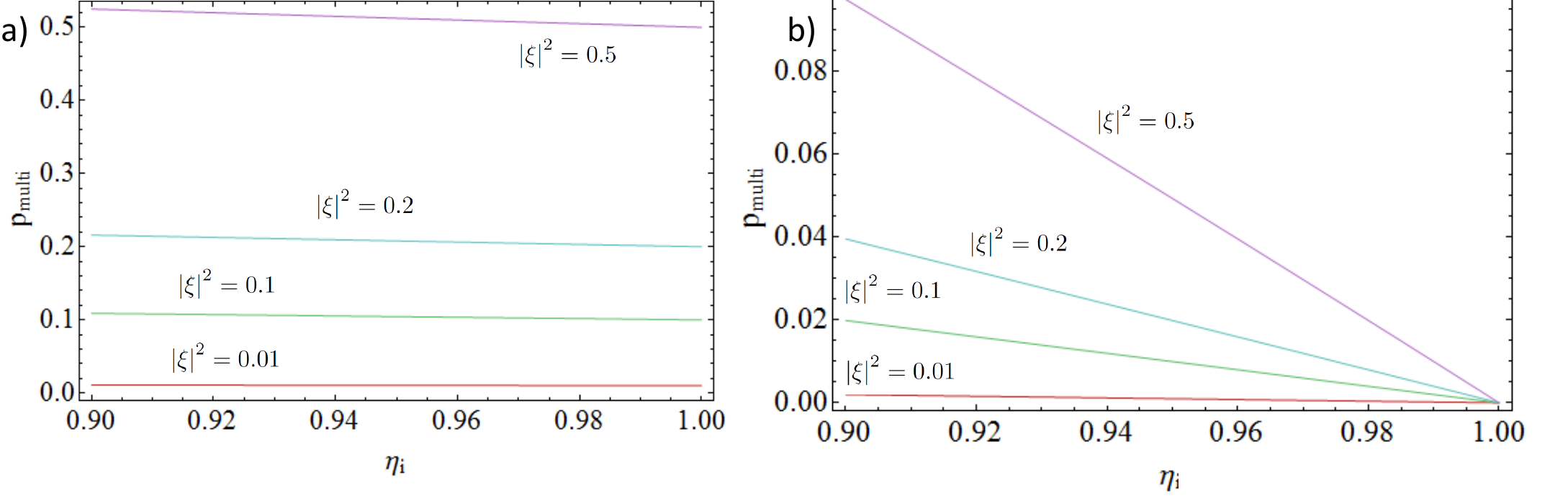}
\par\end{centering}

\protect\caption{Probability of multi-photon emission $p_{multi}$ from a HSPS as a
function of the efficiency in the idler arm $\eta_{i}$ for different
squeezing parameters $\left|\xi\right|^{2}$ using (a) a threshold
detector, (b) a number-resolving detector.\label{fig:HSPScmp}}
\end{figure}

We now have a complete framework for characterizing any HSPS. We assumed
no joint spectral entanglement between photons in a generated
photon pair. However, we can account for joint spectral entanglement by redefining the fidelity
as $p_{single\, D}^{'}=p_{single\, D}\times P$ where $P$ is the
purity of the heralded single photon. Recall that, in the ideal case
of lossless components, the probability $p_{trig\, NRD}$ of heralding
a unit fidelity single photon is bounded by 0.25 (achieved when $\left|\xi\right|^{2}=0.5$
maximizes the probability $\left|\xi\right|^{2}\left(1-\left|\xi\right|^{2}\right)$).
Clearly, HSPS do not suffice on their own to function as a near-deterministic
single-photon source, providing in the ideal scenario one photon every
four pulses on average. 

As we will see in Sec. \ref{sec:Multiplexed-single-photon-sources}, the probability of multi-photon contamination
from individual HSPSs approximates the probability of multi-photon
contamination from MUX sources. We therefore graph $p_{multi\, D}$ as a function
of idler transmission $\eta_{i}$ for several squeezing parameters
$\left|\xi\right|^{2}$ in Figs. \ref{fig:HSPScmp}a and \ref{fig:HSPScmp}b.
To focus on the effect of $\eta_{i}$, we take $\eta_{s}=1$;  the
plots therefore serve as an upper bound on $p_{multi\, D}$. We see
that threshold detectors only reliably herald a single-photon state
for very low levels of squeezing, while number-resolving detectors
 achieve a much lower level of contamination with the same squeezing
parameter. With number-resolving detectors, 10\% loss in the idler
arm ($n_i=0.90$), and pumping with $p_{pair}=0.09$ $(\left|\xi\right|^{2}=0.1)$,
the multi-photon contamination level is almost an order of magnitude
lower ($\sim0.02)$ compared with that of threshold detectors with
the same squeezing parameter ($\sim0.1)$. Despite their enhanced
performance compared to threshold detectors, number-resolving detectors
with the highest pumping parameters are not immune to loss; even 10\%
loss in the idler arm with $p_{pair}=0.25$ $(\left|\xi\right|^{2}=0.5)$
results in a probability of multi-photon contamination of $\sim0.1$.
Achieving a lower level of contamination requires a reduction in either
the squeezing parameter or the loss in the idler arm.

In Sec. \ref{sec:Multiplexed-single-photon-sources} we will show that a practical multiplexed source will
require operation in the strong pumping regime, thus showing that
threshold detectors can not be used for multiplexed sources with the
highest efficiency and low levels of multi-photon contamination. For
this reason, and due to space constraints, we will only consider number-resolving
detectors in the remainder of the paper. However, the derived expression
for threshold detectors can still be used in the framework for multiplexed
sources.

\section{Multiplexed single-photon sources\label{sec:Multiplexed-single-photon-sources}}

\subsection{\label{sub:General-considerations}General considerations}

A general MUX source can be characterised in a similar way to the
HSPS. The probability per clock-cycle  that at least one HSPS in an
array of $N$ HSPSs triggers is given by:

\begin{equation}
p_{trig}^{MUX}=1-\left(1-p_{trig}\right)^{N},\label{eq:ptrigmux}
\end{equation}

and the probability per clock-cycle that at least one source emits
a triggered single-photon is:

\begin{equation}
p_{1}=p_{single}\left(1-\left(1-p_{trig}\right)^{N}\right).\label{eq:MUX_proba_emission}
\end{equation}
$p_{1}$ can in principle be made arbitrarily close to one, such that
a near-deterministic single-photon source (>99\% emission probability) can be made out of 17 HSPSs
using a lossless switching network to route the photon from the HSPS
which triggered to the output \cite{silberhorn_multiplexedPDC_2012}.
However, in any implementation, the switching network will have loss
due to the optical delay lines---required for allowing enough time
to reconfigure the switching network upon trigger from the HSPS---and
the intrinsic loss of the switch components, for example $2\times2$
couplers and phase modulators for MZI-type switches (we will assume
all $2\times2$ switches are MZI-type switches). The network loss,
$\eta_{network}\left(N\right)$ , is proportional to the number of
sources used in the MUX source, since additional sources require a
larger switching network for routing. Provided the losses are equally
distributed in the network (this assumption holds for two of the three
schemes presented in the further sections), the network loss applies
the same amount of loss to every HSPS.

For switching networks with balanced loss, the probability per clock-cycle
for a MUX source to emit a single photon, conditioned on the MUX source
triggering, is given by $p_{single}^{MUX}$, and is calculated from
Eq. \ref{eq:p_single_TD} or \ref{eq:p_single_NRD} (depending on
the type of detector) by replacing $\eta_{s}$ in the expression with
the full transmission $\eta_{s}\eta_{network}(N)$. Similarly, the
probability for a MUX source to emit multi-photon contamination, conditioned
on the MUX source triggering, is given by $p_{multi}^{MUX}$, and
is calculated from either Eq. \ref{eq:p_multi_TD} or \ref{eq:p_multi_NRD}
(depending on the type of detector) by replacing $\eta_{s}$ in the
expression with the full transmission $\eta_{s}\eta_{network}(N)$.
Using this definition of $p_{single}^{MUX}$, it then follows that
the probability per clock-cycle for a multiplexed source to emit a
triggered single-photon can be written:

\begin{equation}
q_{MUX}=p_{single}^{MUX}\times p_{trig}^{MUX}.\label{eq:q_mux_exact}
\end{equation}
For switching networks with balanced loss, a convenient lower bound
for this probability is given by:

\begin{equation}
q_{MUX}^{*}=p_{single}\times p_{trig}^{MUX}\times\eta_{network}(N),\label{eq:q_mux_lower_bound-1}
\end{equation}
which uses only the single-photon emission probability from the HSPSs
and neglects cases in which multi-photon contamination from the HSPS
reduces to a single-photon due to loss from the switching network.

For a multiplexed source pumped by a pulsed laser with repetition
rate $R$, the emission rate of triggered single-photons is:

\[
R_{MUX}=R\times q_{MUX}.
\]

Large-scale linear optical experiments require $M$ multiplexed sources
in parallel for generating $M$ single photons. The $M$-photon state
is heralded by the simultaneous triggering of all $M$ MUX sources
(at least one HSPS per MUX source). For a source pumped by a pulse
laser with repetition rate $R$, the $M$-photon generation rate is:

\begin{equation}
R_{MUX}^{M}=R\times\left(q_{MUX}\right)^{M}.\label{eq:10-photon-rate-single-1}
\end{equation}
The multi-photon contamination probability conditioned on heralding
the M-photon state is:

\begin{equation}
p_{multi}^{MUX\, M}=1-\left(1-p_{multi}^{MUX}\right)^{M}.\label{eq:10-photon-rate-multi-1}
\end{equation}

The analysis so far has been kept general, and can be applied to any
MUX source with balanced network loss. In the next sections we focus
on three specific architectures.

While photon loss and inefficient detectors are likely to be the dominant
source of error for the near-term implementation of MUX sources, we
note that there are other sources of error which will also have an
effect on MUX performance. These include dark counts from single-photon
detectors \cite{rhode-jmo-2006}, mode-mismatch \cite{rhode-pra-2006}, and circuit faults. These
effects are left to a future analysis. We also assume that the detector deadtimes are smaller than the system clock period, and therefore loss effects due to detector deadtimes are not considered in our current work.

\subsection{Log-tree source}

\begin{figure}
\begin{centering}
\includegraphics[scale=0.63]{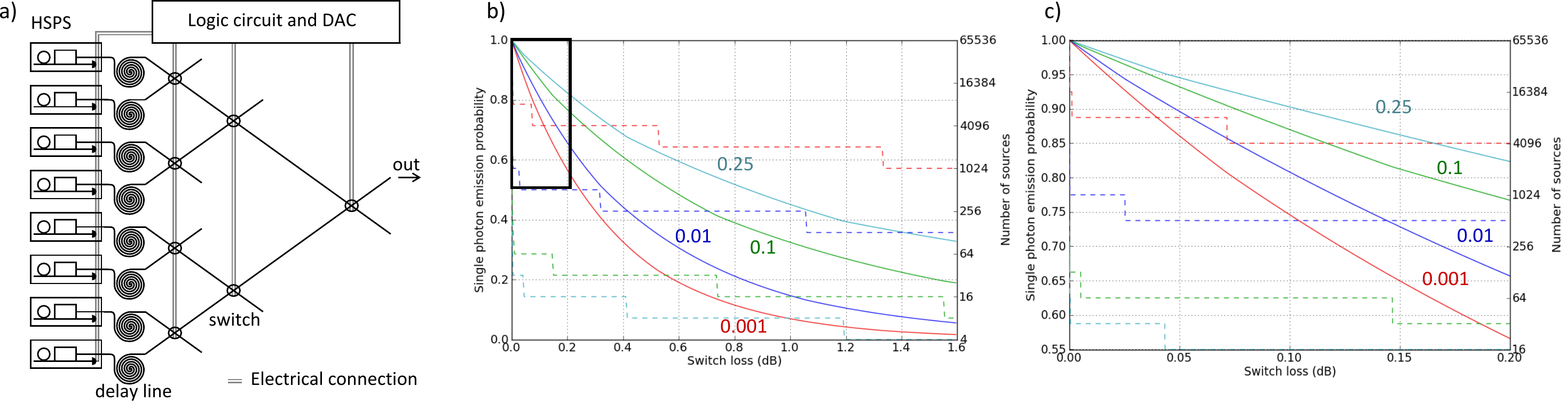}
\par\end{centering}

\protect\caption{\label{fig:Logarithmic-tree}(a) Log-tree MUX source. The trigger
output of each HSPS is linked to logic circuit which configures the
setting of each switch in the network. (b,c) Maximal probability of triggered single-photon emission and optimal number of HSPSs per MUX source as
a function of the switch loss. Each color represents a given
probability to trigger $p_{trig}$.  The plain lines (left axis) represent
the maximal single-photon emission probability for a given switch loss. The dashed line
(right axis) is the  number of HSPSs needed to attain the maximal single-photon emission probability for a given switch loss.  (c) Shows detail at low switch losses (<0.2 dB). We plot the graphs with $\eta_{delay}=1$
and $p_{single}=1$.}
\end{figure}

The first MUX implementation we consider is called the log-tree source.
In this scheme, the output optical ports of $N$ HSPS sources are
connected to an $N$$\times$1 reconfigurable switch, which is a logarithmic
tree composed of $2\times2$ switches (Fig. \ref{fig:Logarithmic-tree}a). Using $N$ sources
requires a log tree with a depth of $\left\lceil \frac{\ln N}{\ln2}\right\rceil $
$2\times2$ switches in order to route any of the $N$ sources to
the output. The signal photons are stored in delay lines as the electrical
trigger output is sent to a logic circuit and the network configuration
is determined and set.

The network loss is given by $\eta_{network}\left(N\right)=\eta^{\left\lceil \frac{\ln N}{\ln2}\right\rceil }\eta_{delay}$
where $\eta$ is the transmission of a switch in the log tree network
and $\eta_{delay}$ is the loss of the delay line. From Eq. \ref{eq:q_mux_exact},
the probability per clock-cycle for the log-tree multiplexed source
to emit a triggered single-photon is given by: 

\begin{equation}
q_{tree}=p_{single}^{tree}\left(1-\left(1-p_{trig}\right)^{N}\right).\label{eq:LogTreeProbaExact}
\end{equation}
and has the lower bound (Eq. \ref{eq:q_mux_lower_bound-1}):

\begin{equation}
q_{tree}^{*}=p_{single}\left(1-\left(1-p_{trig}\right)^{N}\right)\eta^{\left\lceil \frac{\ln N}{\ln2}\right\rceil }\eta_{delay}.\label{eq:LogTreeProba}
\end{equation}
 The optimal number of sources for a given switch loss is found by
numerically finding the $N$ which maximizes Eq. (\ref{eq:LogTreeProba}).
To study the effect of switching loss in isolation from other sources
of loss, the optimal $N$ and $q_{tree}^{*}$ with $\eta_{delay}=1$
and $p_{single}=1$ are plotted as a function of the $2\times2$ switch
loss in Fig. \ref{fig:Logarithmic-tree}b and \ref{fig:Logarithmic-tree}c. As expected, we see that the probability of triggered single-photon
emission tends towards 1 in the limit of low switching loss for all
trigger probabilities. The number of HSPSs required in the weak pump
regime ($p_{trig}$ < 0.01) is likely to be impractical: obtaining
a probability of single photon emission $q_{tree}^{*}>0.9$ with $p_{trig}=0.01$
requires 512 sources in parallel and switches with $0.05$~dB loss ($\sim0.989$ transmission).
Obtaining  $q_{tree}^{*}>0.9$ with $p_{trig}=0.1$ requires 64 sources
in parallel and switches with $0.07$~dB loss ($\sim0.984$ transmission). With the maximum trigger
probability $p_{trig}=0.25$, only 16 sources and switches with 0.1~dB
loss ($\sim0.977$ transmission) are required.

Because the switching network is balanced, the probability for a triggered
log-tree MUX source to emit multi-photon contamination $p_{multi}^{MUX}$
is calculated as explained in Sec. \ref{sub:General-considerations}
using $\eta_{network}\left(N\right)=\eta^{\left\lceil \frac{\ln N}{\ln2}\right\rceil }\eta_{delay}$.
Since the loss from the switching network can only decrease $p_{multi}^{MUX}$,
the multi-photon contamination probabilities for the HSPSs in Figs.
\ref{fig:HSPScmp}a) and b) serve as valid upper bounds for $p_{multi}^{MUX}$.
We will consider these expressions further when we consider $M$-photon
state generation in Sec. \ref{sec:10-photon-state-generation}.

\subsection{Generalized Mach-Zehnder interferometer}

\begin{figure}[H]
\begin{centering}
\includegraphics[scale=0.7]{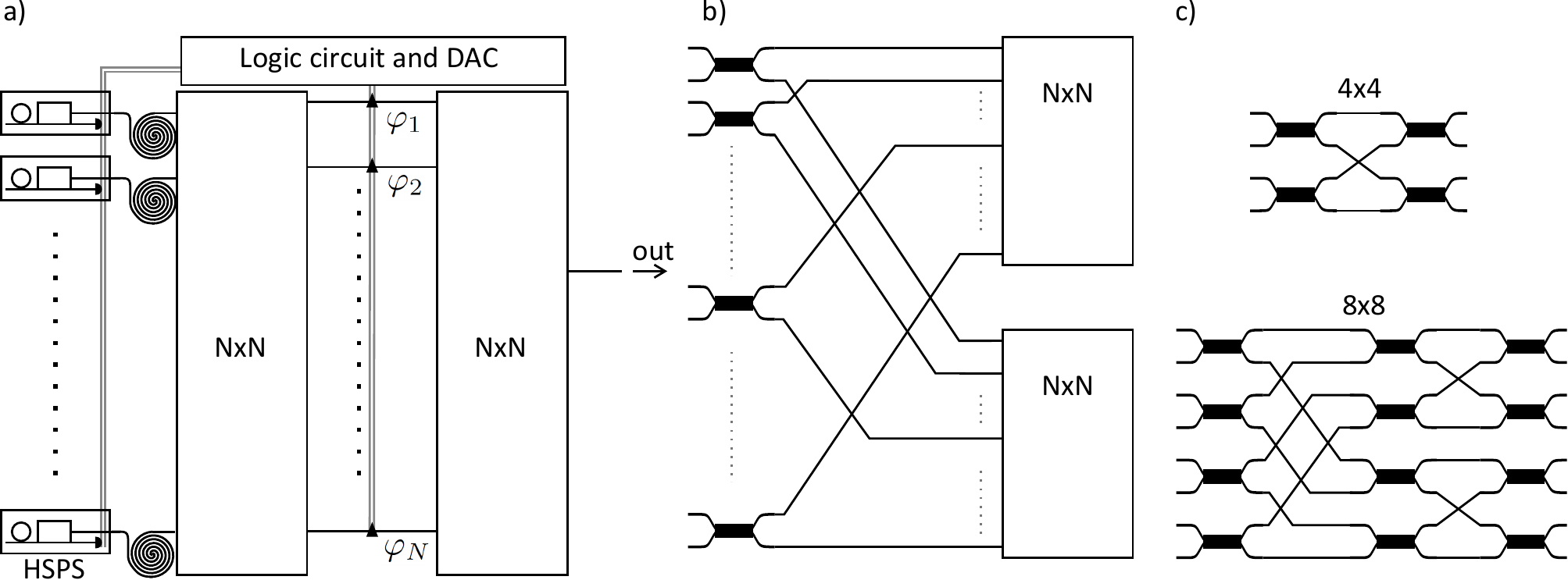}
\par\end{centering}

\protect\caption{\label{fig:Generalized-Mach-Zehnder-for}Generalized Mach-Zehnder
MUX source. a) Multiplexing $N$ HSPSs using a generalized MZI composed
of two $N\times N$ balanced splitters enclosing $N$ arms each having
a tuneable phase $\varphi_{i}$. b) Recursive definition of the $N\times N$
balanced coupler. A $2N\times2N$ balanced coupler is composed of
$N$ 2 \texttimes{} 2 couplers each having one arm connected to a
$N\times N$ balanced coupler and the other arm connected to another
$N\times N$ balanced coupler. c) Examples of circuit layouts for
the 4 \texttimes{} 4 and 8 \texttimes{} 8 balanced couplers.}
\end{figure}

\begin{figure}[H]
\begin{centering}
\includegraphics[scale=0.89]{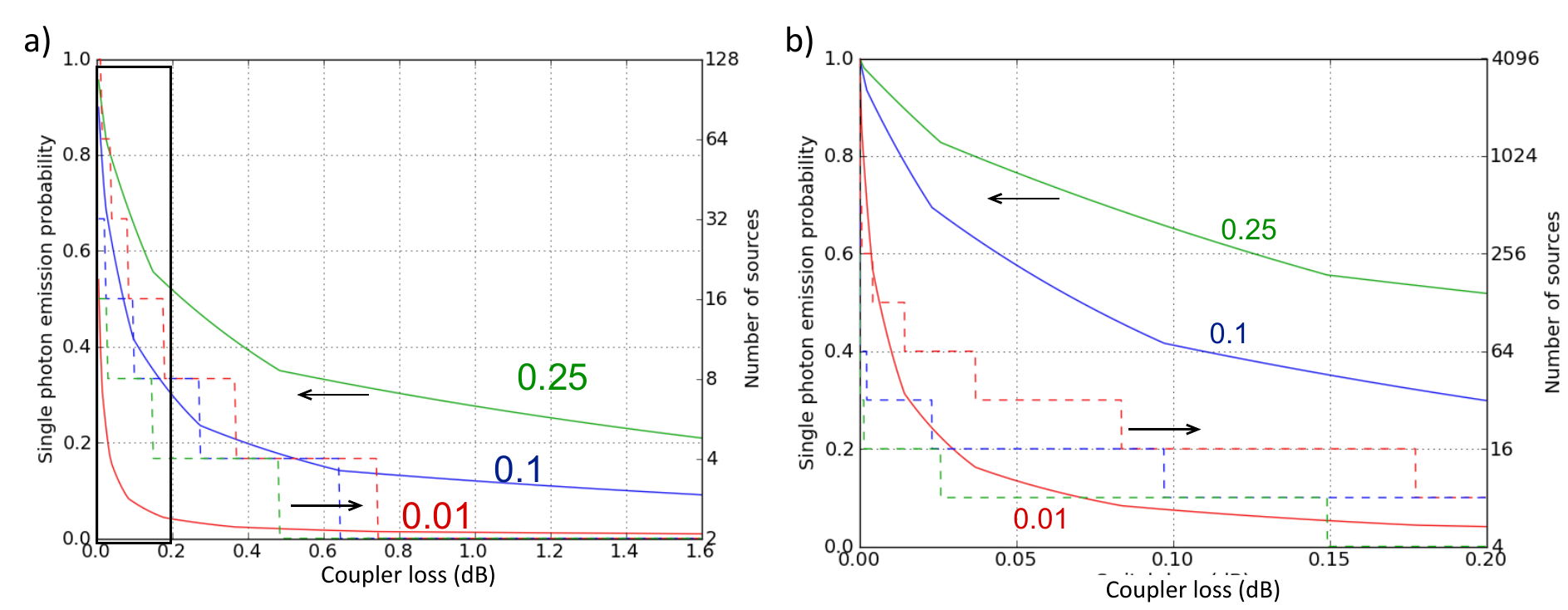}
\par\end{centering}

\protect\caption{\label{fig:Single-photon-probability}GMZ MUX source: Maximal probability of triggered single-photon emission and optimal number of HSPSs per MUX source as
a function of the switch loss. Each color represents a given
probability to trigger $p_{trig}$.  The plain lines (left axis) represent
the maximal single-photon emission probability for a given coupler loss. The dashed line
(right axis) is the number of HSPSs needed to attain the maximal single-photon emission probability for a given coupler loss.  (c) Shows detail at low coupler losses (<0.2 dB). We plot the graphs with $\eta_{delay}$,$\eta_{modulator}$,
and $p_{single}=1$.}
\end{figure}

Like the log-tree source, this scheme also uses $N$ HSPS sources
connected to a $N\times1$ reconfigurable switch (Fig. \ref{fig:Generalized-Mach-Zehnder-for}a).
However, here the $N\times1$ switch is a generalized Mach-Zehnder
interferometer (GMZ) - composed of two $N\times N$ balanced splitters
enclosing $N$ phase modulators. $N$ phase modulators are sufficient
to route any input to a given fixed output port. This is achieved
by setting half of the phases to $\pi$ and setting the other half
to 0, or by applying 0 to all the phases to obtain a full swap. The
$N\times N$ passive splitter can either be a $N\times N$ MMI or
built from of 2$\times$2 couplers. Fabricating large $N\times N$
balanced MMIs with low loss is challenging, so we propose using cascaded
couplers (having a reflectivity of 0.5) and crossings as shown in
Figs. \ref{fig:Generalized-Mach-Zehnder-for}b and \ref{fig:Generalized-Mach-Zehnder-for}c.

The lower bound on the probability to emit a triggered single photon
$q_{GMZ}^{*}$ is, from Eq. \ref{eq:q_mux_lower_bound-1}:

\begin{equation}
q_{GMZ}^{*}=p_{single}\eta_{delay}\left(1-\left(1-p_{trig}\right)^{N}\right)\eta_{modulator}\eta_{N\times N}^{2},\label{eq:gmzprob}
\end{equation}
where $\eta_{modulator}$ is the transmission of the modulator section
and $\eta_{N\times N}$ is the loss induced by the balanced $N\times N$
switch. If implemented with couplers with transmission $\eta_{coupler}$,
then $\eta_{N\times N}=\eta_{coupler}^{N-1}$. To show the effect
of coupler loss, the optimal number of sources to achieve a probability
of triggered single photon emission $q_{GMZ}^{*}$ with $p_{single}\eta_{delay}\eta_{modulator}=1$
is plotted in Fig. \ref{fig:Single-photon-probability}.

The choice between the log tree and the GMZ depends on the dominant
component loss. In the GMZ, the photon must pass only a single phase
modulator, instead of a logarithmic number as in the log-tree scheme.
However, the GMZ requires a linear scaling in the number of directional
couplers the photon passes through, instead of the logarithmic scaling
given by the log tree scheme. Also, the GMZ requires $O(n^{2})$ couplers
while the total number of components for the log tree scales linearly.
Table (1) shows a summary of the different resource scalings for the
two architectures. 

Since the switching network is balanced, the exact probability for
emitting a triggered single-photon $q_{GMZ}$ and the output state
multi-photon contamination probability $p_{multi}^{GMZ}$ are calculated
as explained in section (\ref{sub:General-considerations}) using
$\eta_{network}=\eta_{delay}\eta_{modulator}\eta_{N\times N}^{2}$.
We will consider these expressions further in section \ref{sec:10-photon-state-generation}.

\begin{table}
\begin{tabular}{|>{\centering}p{3cm}|>{\centering}p{3cm}|>{\centering}p{3cm}|>{\centering}p{3cm}|>{\centering}p{3cm}|}
\hline 
Source Type & Total number of phase modulators & Total number of directional couplers & Phase modulator depth & Coupler depth\tabularnewline
\hline 
\hline 
Log Tree & $N-1$ & 2$\left(N-1\right)$ & $\log_{2}\left(N\right)$ & 2$\log_{2}\left(N\right)$\tabularnewline
\hline 
GMZ & $N$ & $\frac{N\left(N+\log_{2}\left(N\right)-1\right)}{4}$ & 1 & 2$\left(N-1\right)$\tabularnewline
\hline 
\end{tabular}\protect\caption{Comparison of the requirements and component depth between the log
tree and generalized MZI architectures as a function of the number
of HSPSs $N$. The first two columns give the total number of component
of the given type required for building the switch. The last two columns
provide the depth of the circuit for each component---or the number
of components of each type that each photon has to go through.}
\end{table}

\subsection{Chained sources}

\begin{figure}
\begin{centering}
\includegraphics[scale=0.55]{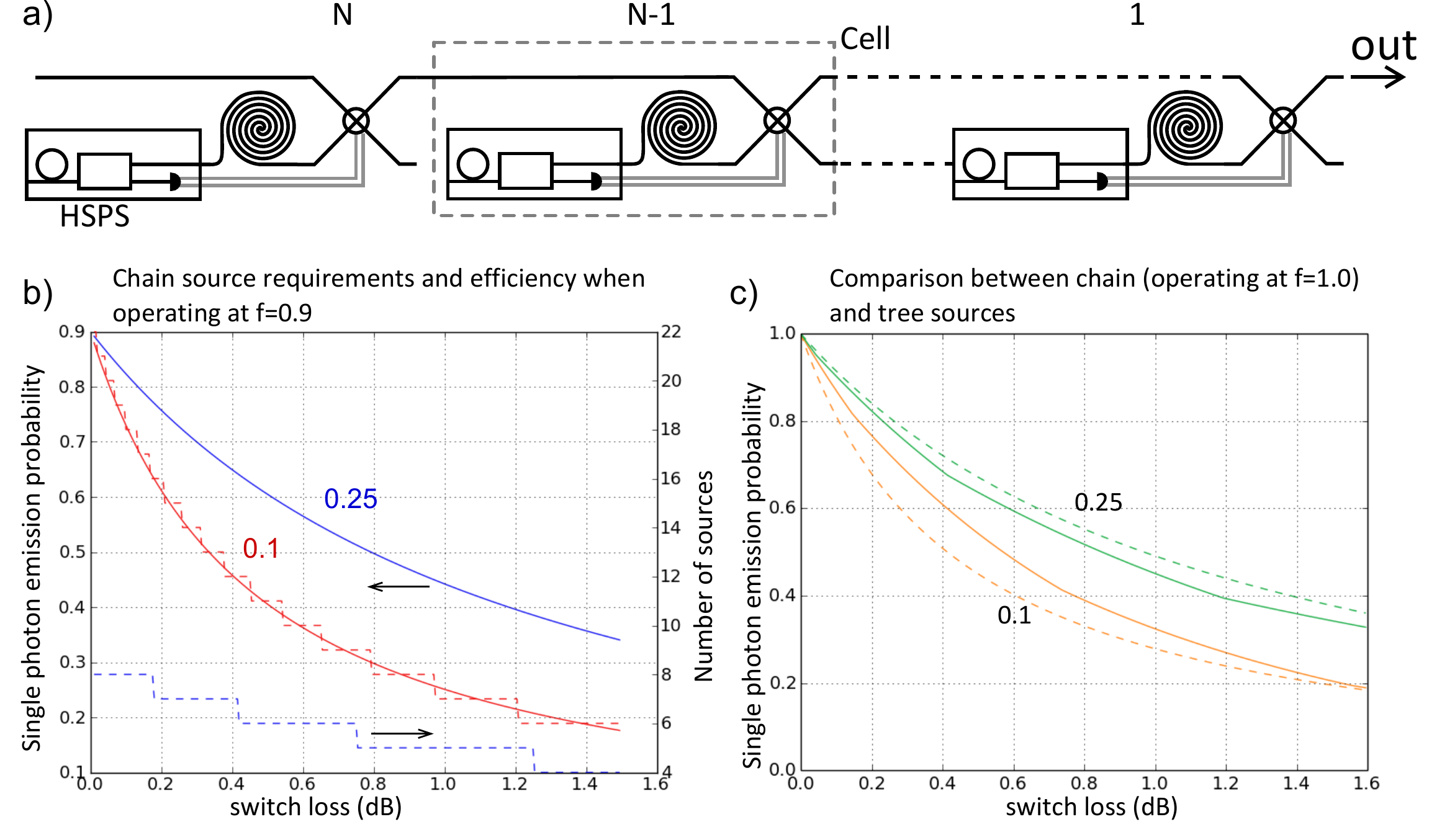}
\par\end{centering}

\protect\caption{\label{fig:Chain-source-scheme}Chain MUX source and performance.
a) Figure 6: Array of chained sources. N unit cells are cascaded to
build a chained multiplexed single-photon source. A unit cell is composed
of a HSPS, an optical delay line, and a 2 \texttimes{} 2 switch. b)
Number of cells required (dash line, right axis) to achieve a maximal
probability of triggered single-photon emission $q_{chain}^{*}$ (plain
line, left axis) as a function of the switch loss. We chose a fractional efficiency of  f = 0.9 
 to compute the required number of cells. Two regimes are plotted corresponding
to $p_{trig}=0.1$ (red) and $p_{trig}=0.25$ (blue). c) Comparison
between the chained source (plain) and log tree (dashed) triggered single-photon emission probability, as a function of switch loss, for two different regimes,
$p_{trig}=0.1$ (orange) and $p_{trig}=0.25$ (green). The optimal single-photon probability $q_{max}^{*}$
is plotted for both the chain source---for f=1.0, corresponding to the limit of an infinite number of cascaded cells---and for the log-tree source---using the optimal number of switches.
}
\end{figure}

In this scheme, unit cells---each composed of a HSPS, a delay line,
and a $2\times2$ switch---are cascaded to build a chain of sources
(Fig. \ref{fig:Chain-source-scheme}a). The electrical output trigger
of the HSPS directly drives a switch which routes the emitted photon
to the output port while the input port is routed towards a blocked
path. As in the other schemes, an optical delay line allows sufficient
time for the detection of the idler photon and the configuration of
the switch.

The chained scheme benefits from very simple control logic requirements,
since each switch is only driven by one HSPS. For switches driven
by voltage, for example, the logic consists only of amplifying the
trigger signal from the HSPS to the required switching voltage. This
switching logic  privileges the HSPS emitting closest to the output
of the chain, and hence the photon from the HSPS suffering the least
from the losses. As an example, if two cells are cascaded and cell
2 triggers before cell 1, the switch corresponding to cell 1 will
remove the photon from the HSPS of cell 2 and replace it with a new
photon from the HSPS of cell 1.

Since the switching network applies different amounts of loss to the
different HSPSs, we cannot derive the lower bound for the single-photon
emission probability $q_{chain}^{*}$ using the method given in Sec.
\ref{sub:General-considerations}. We therefore show a different derivation
here. Given a chain of $N$ unit cells, assuming HSPSs with a probability
of triggering $p_{trig}$, a probability for the heralded state to
be a single photon $p_{single}$, and switches with transmission $\eta$,
$q_{chain}^{*}$ is given by summing the probabilities for each source
to fire given that the subsequent ones do not, and accounting for
switching losses, which gives:

\[
q_{chain}^{*}=p_{single}\eta_{delay}\sum_{n=1}^{N}p_{trig}\eta^{n}\left(1-p_{trig}\right)^{n-1}
\]

\begin{equation}
=p_{single}\eta_{delay}p_{trig}\eta\frac{1-\left[\left(1-p_{trig}\right)\eta\right]^{N}}{1-\left(1-p_{trig}\right)\eta}.\label{eq:chain}
\end{equation}

The maximal probability for emitting a single photon is obtained in
the limit of an infinite chain $q_{max}=p_{single}\eta_{delay}\frac{p_{trig}\eta}{1-\left(1-p_{trig}\right)\eta}$.
In practice, to operate at a fraction $f$ of the maximal probability
$q_{max}^{*}$, the length of the chain is given by solving $fq_{max}^{*}=p_{single}\eta_{delay}p_{trig}\eta\frac{1-\left[\left(1-p_{trig}\right)\eta\right]^{N}}{1-\left(1-p_{trig}\right)\eta}$
which gives:

\begin{equation}
N=\left\lceil \frac{\ln\left(1-x\right)}{\ln\left[\left(1-p_{trig}\right)\eta\right]}\right\rceil .\label{eq:N}
\end{equation}

We show the single-photon emission probability $q_{chain}^{*}$ with $p_{single}\eta_{delay}=1$
and the number of cells required $N$ as a function of the switching
loss for $f=0.9$ in Fig. \ref{fig:Chain-source-scheme}b. As with
the log tree architecture, the required number of sources becomes
impractical for small $p_{trig}$, but can be kept below 8 when operating
at $p_{trig}=0.25$. The performance of the chained scheme compares
well with the performance of the log tree, as shown in \ref{fig:Chain-source-scheme}c).
It has better performances for $p_{trig}=0.1$ (for a switch loss
less than 1.6\,dB) and is close to the log-tree performances for
$p_{trig}=0.25$.

The derivation for the exact probability of triggered single photon
emission $q_{chain}$ and the output state multi-photon contamination
probability $p_{multi}^{chain}$ are shown in Appendix B.

\section{Discussion: Generating M Single Photons Using M Multiplexed Sources\label{sec:10-photon-state-generation}}

\begin{figure}
\begin{centering}
\includegraphics[scale=0.4]{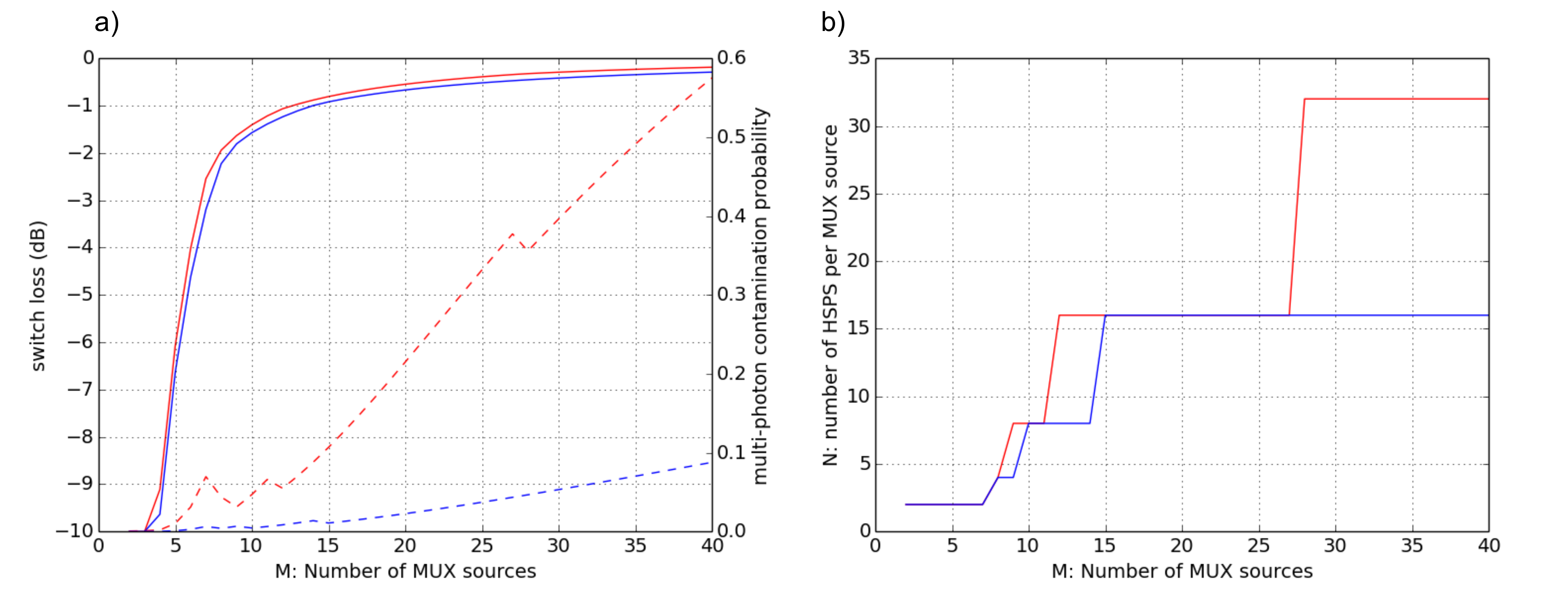}
\par\end{centering}

\protect\caption{Left axis (plain lines):Requirements for generating M single photons
at a rate of 100 Hz, from a 100 MHz pulsed laser for two different idler detection efficiencies. The red lines
correspond to a lumped detection efficiency of
$\eta_{i}=\eta_{s}=0.9$. The blue lines correspond to a lumped detection efficiency of $\eta_{i}=\eta_{s}=0.99$.
a) Maximum tolerable switch loss (per 2x2 switch) vs number M of log-tree MUX sources operated in parallel. Right axis (dashed lines): probability
of multi-photon contamination in output state from the M MUX sources when using the maximum tolerable switch loss.  
b)\label{fig:SwitchLossVsNbMUX-1}  Number of HSPSs required per MUX source in order to meet the requirements in a).
}
\end{figure}

Experiments with about 8 single photons represent the current state-of-the-art
photon number  for SPDC experiments in practice, so we will examine
the performance of multiplexed sources for the generation of  >10 single photons in separate spatial modes.
We first consider using $M$ log-tree multiplexed sources to generate
$M$ single photons at a rate of 100 Hz, assuming a pulsed
laser seed with a repetition rate of 100 MHz, number-resolving detectors,
and $p_{pair}=0.1$. Fig.  \ref{fig:SwitchLossVsNbMUX-1} shows the
maximum tolerable switch loss for generating $M$ single photons under these assumptions, and the resulting probability of multi-photon
contamination in the output state for this same switch loss. We considered two values for the lumped idler detection efficiency (which includes filter loss), $\eta_{i}=0.9$ and $\eta_{i}=0.99$, and assumed that this loss is also applied to the signal channel. We see that with $\eta_{i}=0.9$ and switches with $1$~dB  loss ($\sim0.794$ transmission), $14$ single photons can be generated at the target 100 Hz rate
with multi-photon contamination  below $10\%$ 
. Extending beyond this regime to perform
experiments with 20-40 photons with less than 10\% multi-photon contamination
 requires $\eta_{i}\approx0.99$ lumped idler detection efficiency
with 2x2 switches having each $<0.4-0.2$ dB loss ($\sim0.912-0.955$ transmission).  

Although these numbers are beyond what is attainable using the current state-of-the-art components, they may be attainable in the near-to-mid-future with further development and integration. For reference, prototype all-optical switches in fiber have demonstrated insertion
loss of $\sim0.6$ dB ($\sim0.871$ transmission) \cite{rambo2013low}, on-chip filters for sufficient pump suppression have demonstrated insertion loss of of $\sim0.8$ dB ($\sim0.832$ transmission) \cite{Jeong:13}, and state-of-the-art single-photon
detectors have demonstrated efficiencies of 0.93 \cite{MarsiliF.:2013aa}.  Although the single-photon detectors referenced in \cite{MarsiliF.:2013aa} are non-number-resolving and are limited to a deadtime of 40 ns, we note that passive multiplexing techniques using arrays of detectors \cite{PhysRevLett.76.2464}\cite{Sahin2013} can achieve approximate number-resolving capabilities and reduce deadtime.  However, the exact effect of loss using these specific detector architectures on multiplexed sources, and their resource requirements, is beyond the scope of this paper. 

For applications requiring a very large number of single-photons ($M\gg40$),
such as the universal quantum computer, error-correcting and loss-tolerant
encodings can be used to achieve fault-tolerant computation, provided
the error rates are below required thresholds. Meeting these thresholds
will require sources with the highest single-photon emission probability
and very low multi-photon contamination. We briefly consider what
is attainable with optimistic long-term goals for the performance
of required hardware components. 
Assuming 0.99 lumped idler detection efficiency, 2x2 switches with
0.98 transmission, and $p_{pair}=0.1$, this corresponds to a multiplexed single-photon
source efficiency of 0.8744, a multi-photon emission probability of
0.0017, and 64 HSPSs per MUX source. Using $p_{pair}=0.25$, this
increases the single-photon source efficiency to 0.8965 and reduces
the resource requirements to 16 HSPS per MUX source, although at the
cost of an increased multi-photon contamination probability of 0.0083.
Further research is necessary to translate these single- and multi-photon
emission probabilities into error rates for fault-tolerance in linear
optical quantum computing, and to optimize multiplexed source architectures
for scalable quantum computation.

~

~

~

\section{Conclusion}

In our analysis we have studied multiplexed single-photon sources
with photon loss and inefficient detectors. We derived expressions
for single-photon emission probability and multi-photon contamination
probability using number-resolving and non-number-resolving detectors
for a general HSPS and for three MUX architectures: a log-tree switching
scheme, a chained switching scheme, and a generalized MZI scheme. 

Our findings indicate that number-resolving detectors offer a considerable
advantage over threshold detectors for MUX sources, and are essential
for MUX sources with the highest efficiency. The performance of the
chained scheme compares well with the performance of the log tree,
and has a simple logic requirements that may be beneficial for a near-term
implementation. The GMZ offers an alternative switching structure
that requires fewer phase modulators but more directional couplers
than the other schemes. 

All three of the architectures considered are capable of a high efficiency
single-photon source with low multi-photon contamination if the components
are close to ideal. For example, with a log-tree architecture composed
of switches with $<0.4-0.2$ dB loss ($\sim0.912-0.955$ transmission) and number-resolving
detectors with 99\%  efficiency, a single-photon source capable of producing a
100 Hz rate of 20-40 photon states with less than 10\% multi-photon
contamination is possible. Such a source would be a valuable resource
for quantum technologies such as the boson sampling machine and quantum
simulators. With lower-loss switches and the high efficiency detectors,
a MUX source approaching the threshold requirements for a fully fault-tolerant
universal quantum computer should be possible. Further work is necessary
to find the optimal multiplexing schemes which could encompass the
direct generation and multiplexing of multi-photon states or entangled
resource states. 

~

We thank Xiao-Qi Zhou, Joshua Silverstone, Jonathan Matthews, Anthony
Laing, Daryl Beggs, and Jake Kennard for useful discussions and comments.  This work was supported by Army Research Office (ARO) grant
No. W911NF-14-1-0133.

\section*{Appendix A: Heralded single-photon source parameters derivation}

\subsubsection*{Threshold detector}

The source output state after accounting for losses is given by: 
\begin{equation}
\noindent\hat{\rho}=\left(1-\left|\xi\right|^{2}\right)\left(\sum_{n=0}^{\infty}\left|\xi\right|^{2n}\sum_{p=0}^{n}\sum_{k=0}^{n}C_{n}^{p}\eta_{i}^{p}\left(1-\eta_{i}\right)^{n-p}C_{n}^{k}\eta_{s}^{k}\left(1-\eta_{s}\right)^{n-k}\hat{\rho}_{p,k}\right),\label{eq:pair genration reduced state-1}
\end{equation}
where $\hat{\rho}_{p,k}=\left|p\right\rangle _{i}\left|k\right\rangle _{s}\left\langle p\right|_{i}\left\langle k\right|_{s}$
and $C_{n}^{k}$ is the binomial coefficient.

The probability for the detector placed on the idler arm to trigger
is given by summing all the contribution of states having at least
one photon. Starting with the reduced state after tracing out the
signal mode:

\[
\noindent p_{trig\, TD}=\sum_{m=1}^{\infty}\left\langle m\right|_{i}\left(1-\left|\xi\right|^{2}\right)\left(\sum_{n=0}^{\infty}\left|\xi\right|^{2n}\sum_{p=0}^{n}C_{n}^{p}\eta_{i}^{p}\left(1-\eta_{i}\right)^{n-p}\left|p\right\rangle _{i}\left\langle p\right|_{i}\right)\left|m\right\rangle _{i}
\]

\[
\noindent=\left(1-\left|\xi\right|^{2}\right)\left(\sum_{n=1}^{\infty}\left|\xi\right|^{2n}\sum_{p=1}^{n}C_{n}^{p}\eta_{i}^{p}\left(1-\eta_{i}\right)^{n-p}\right)
\]

\[
\noindent=\left(1-\left|\xi\right|^{2}\right)\left(\sum_{n=1}^{\infty}\left|\xi\right|^{2n}\left(\sum_{p=0}^{n}C_{n}^{p}\eta_{i}^{p}\left(1-\eta_{i}\right)^{n-p}-\left(1-\eta_{i}\right)^{n}\right)\right)
\]

\[
=\left(1-\left|\xi\right|^{2}\right)\left(\sum_{n=0}^{\infty}\left|\xi\right|^{2n}\left(1-\left(1-\eta_{i}\right)^{n}\right)\right)
\]

\[
=\left(1-\left|\xi\right|^{2}\right)\left(\frac{1}{1-\left|\xi\right|^{2}}-\frac{1}{1-\left|\xi\right|^{2}\left(1-\eta_{i}\right)}\right).
\]

\begin{equation}
p_{trig\, TD}=\frac{\left|\xi\right|^{2}\eta_{i}}{1-\left|\xi\right|^{2}\left(1-\eta_{i}\right)}.\label{eq:trig_TD}
\end{equation}

We can check the validity of the expression in the two extreme cases.
If $\eta_{i}=1$, the probability to trigger is maximal, $p_{trig}=\left|\xi\right|^{2}$.
If $\eta_{i}=0$ then $p_{trig}=0$.

The heralded state in the signal arm is expressed, renormalizing by
dividing by $p_{trig}$, as:

\[
\hat{\rho}_{H\, TD}=\frac{\left(1-\left|\xi\right|^{2}\right)}{p_{trig\, TD}}\sum_{m=1}^{\infty}\left\langle m\right|_{i}\left(\sum_{n=0}^{\infty}\left|\xi\right|^{2n}\sum_{p=0}^{n}\sum_{k=0}^{n}C_{n}^{p}\eta_{i}^{p}\left(1-\eta_{i}\right)^{n-p}C_{n}^{k}\eta_{s}^{k}\left(1-\eta_{s}\right)^{n-k}\hat{\rho}_{p,k}\right)\left|m\right\rangle _{i}
\]

\[
=\frac{\left(1-\left|\xi\right|^{2}\right)}{p_{trig\, TD}}\left(\sum_{n=1}^{\infty}\left|\xi\right|^{2n}\sum_{p=1}^{n}\sum_{k=0}^{n}C_{n}^{p}\eta_{i}^{p}\left(1-\eta_{i}\right)^{n-p}C_{n}^{k}\eta_{s}^{k}\left(1-\eta_{s}\right)^{n-k}\left|k\right\rangle _{s}\left\langle k\right|_{s}\right).
\]

We can compute the probability that the heralded state is a single photon
using $p_{single\, TD}=\left\langle 1\right|_{s}\hat{\rho}_{heralded\, TD}\left|1\right\rangle _{s}$:

\[
\noindent p_{single\, TD}=\frac{\left(1-\left|\xi\right|^{2}\right)}{p_{trig\, TD}}\left(\sum_{n=1}^{\infty}\left|\xi\right|^{2n}C_{n}^{1}\eta_{s}\left(1-\eta_{s}\right)^{n-1}\sum_{p=1}^{n}C_{n}^{p}\eta_{i}^{p}\left(1-\eta_{i}\right)^{n-p}\right)
\]

\[
=\frac{\left(1-\left|\xi\right|^{2}\right)}{p_{trig\, TD}}\left(\sum_{n=1}^{\infty}\left|\xi\right|^{2n}C_{n}^{1}\eta_{s}\left(1-\eta_{s}\right)^{n-1}\left(1-\left(1-\eta_{i}\right)^{n}\right)\right)
\]

\[
=\frac{\left(1-\left|\xi\right|^{2}\right)}{p_{trig\, TD}}\left|\xi\right|^{2}\left(\eta_{s}\sum_{n=1}^{\infty}n\left|\xi\right|^{2(n-1)}\left(1-\eta_{s}\right)^{n-1}-\left(1-\eta_{i}\right)\eta_{s}\sum_{n=1}^{\infty}\left|\xi\right|^{2(n-1)}n\left(1-\eta_{s}\right)^{n-1}\left(1-\eta_{i}\right)^{n-1}\right).
\]

Using the identity (obtained from taking the derivative of the geometric
series) $\sum_{n=0}^{\infty}nx^{n-1}=1/(1-x)^{2}$:

\[
p_{single\, TD}=\frac{\left(1-\left|\xi\right|^{2}\right)}{p_{trig\, TD}}\left|\xi\right|^{2}\eta_{s}\left(\frac{1}{\left(1-\left|\xi\right|^{2}\left(1-\eta_{s}\right)\right)^{2}}-\frac{\left(1-\eta_{i}\right)}{\left(1-\left|\xi\right|^{2}\left(1-\eta_{s}\right)\left(1-\eta_{i}\right)\right)^{2}}\right).
\]

\begin{equation}
\noindent p_{single\, TD}=\left(1-\left|\xi\right|^{2}\right)\eta_{s}\frac{\left[1-\left(\left|\xi\right|^{2}\left(1-\eta_{s}\right)\right)^{2}\left(1-\eta_{i}\right)\right]\left[1-\left|\xi\right|^{2}\left(1-\eta_{i}\right)\right]}{\left[1-\left|\xi\right|^{2}\left(1-\eta_{s}\right)\right]^{2}\left[1-\left|\xi\right|^{2}\left(1-\eta_{s}\right)\left(1-\eta_{i}\right)\right]^{2}.}\label{eq:p_single_TD-1}
\end{equation}

In the lossless case, we obtain $p_{single\, TD}=\left(1-\left|\xi\right|^{2}\right)$,
which, with a triggering probability $p_{trig}=\left|\xi\right|^{2}$
(from Eq. \ref{eq:trig_TD}), is consistent with a probability of
getting one photon of $\left|\xi\right|^{2}\left(1-\left|\xi\right|^{2}\right)$
(from Eq. \ref{eq:SPDC state}).

The probability that the heralded state contains multi-photon contamination
is derived as follows. First, we define:

\[
Z_{TD}=\sum_{q=1}^{\infty}\sum_{m=1}^{\infty}\left\langle q,m\right|\hat{\rho}\left|q,m\right\rangle 
\]

\[
=\left(1-\left|\xi\right|^{2}\right)\left(\sum_{n=1}^{\infty}\left|\xi\right|^{2n}\sum_{p=1}^{n}\sum_{k=1}^{n}C_{n}^{p}\eta_{i}^{p}\left(1-\eta_{i}\right)^{n-p}C_{n}^{k}\eta_{s}^{k}\left(1-\eta_{s}\right)^{n-k}\right)
\]

\[
=\left(1-\left|\xi\right|^{2}\right)\left(\sum_{n=1}^{\infty}\left|\xi\right|^{2n}\left(1-\left(1-\eta_{i}\right)^{n}\right)\left(1-\left(1-\eta_{s}\right)^{n}\right)\right)
\]

\[
=\left(1-\left|\xi\right|^{2}\right)\left(\sum_{n=1}^{\infty}\left|\xi\right|^{2n}\left(1+\left(1-\eta_{s}\right)^{n}\left(1-\eta_{i}\right)^{n}-\left(1-\eta_{i}\right)^{n}-\left(1-\eta_{s}\right)^{n}\right)\right)
\]

\[
=\left(1-\left|\xi\right|^{2}\right)\left|\xi\right|^{2}\left(\frac{1}{1-\left|\xi\right|^{2}}+\frac{\left(1-\eta_{s}\right)\left(1-\eta_{i}\right)}{1-\left|\xi\right|^{2}\left(1-\eta_{s}\right)\left(1-\eta_{i}\right)}-\frac{\left(1-\eta_{i}\right)}{1-\left|\xi\right|^{2}\left(1-\eta_{i}\right)}-\frac{\left(1-\eta_{s}\right)}{1-\left|\xi\right|^{2}\left(1-\eta_{s}\right)}\right).
\]

Then it follows that $p_{multi\, TD}=\sum_{m=2}^{\infty}\left\langle m\right|_{s}\hat{\rho}_{heralded\, TD}\left|m\right\rangle _{s}$:

\begin{equation}
p_{multi\, TD}=\frac{Z_{TD}}{p_{trig\, TD}}-p_{single\, TD}.\label{eq:p_multi_TD-1}
\end{equation}

\subsubsection*{Number-resolving detector}

Now we consider the number-resolving detector. Starting with Eq. \ref{eq:pair genration reduced state},
tracing out the signal mode and calculating the probability to get
only one photon:

\[
p_{trig\, NRD}=\left\langle 1\right|_{i}\left(1-\left|\xi\right|^{2}\right)\left(\sum_{n=0}^{\infty}\left|\xi\right|^{2n}\sum_{p=0}^{n}C_{n}^{p}\eta_{i}^{p}\left(1-\eta_{i}\right)^{n-p}\left|p\right\rangle _{i}\left\langle p\right|_{i}\right)\left|1\right\rangle _{i}
\]

\[
=\left(1-\left|\xi\right|^{2}\right)\left(\sum_{n=1}^{\infty}\left|\xi\right|^{2n}C_{n}^{1}\eta_{i}\left(1-\eta_{i}\right)^{n-1}\right)
\]

\[
=\left(1-\left|\xi\right|^{2}\right)\left|\xi\right|^{2}\eta_{i}\left(\sum_{n=1}^{\infty}\left|\xi\right|^{2\left(n-1\right)}n\left(1-\eta_{i}\right)^{n-1}\right).
\]

\begin{equation}
p_{trig\, NRD}=\frac{\left(1-\left|\xi\right|^{2}\right)\left|\xi\right|^{2}\eta_{i}}{\left(1-\left(1-\eta_{i}\right)\left|\xi\right|^{2}\right)^{2}}.
\end{equation}

We can check the validity of the expression in the two extreme cases.
If $\eta_{i}=1$, the probability to trigger is maximal, $p_{trig}=\left(1-\left|\xi\right|^{2}\right)\left|\xi\right|^{2}$.
If $\eta_{i}$=0 then $p_{trig}=0$.

The corresponding heralded state is:

\[
\hat{\rho}_{H\, NRD}=\frac{\left(1-\left|\xi\right|^{2}\right)}{p_{trig\, NRD}}\left\langle 1\right|_{i}\left(\sum_{n=0}^{\infty}\left|\xi\right|^{2n}\sum_{p=0}^{n}\sum_{k=0}^{n}C_{n}^{p}\eta_{i}^{p}\left(1-\eta_{i}\right)^{n-p}C_{n}^{k}\eta_{s}^{k}\left(1-\eta_{s}\right)^{n-k}\hat{\rho}_{p,k}\right)\left|1\right\rangle _{i}
\]

\[
=\frac{\left(1-\left|\xi\right|^{2}\right)}{p_{trig\, NRD}}\eta_{i}\left(\sum_{n=1}^{\infty}\left|\xi\right|^{2n}n\left(1-\eta_{i}\right)^{n-1}\sum_{k=0}^{n}C_{n}^{k}\eta_{s}^{k}\left(1-\eta_{s}\right)^{n-k}\left|k\right\rangle _{s}\left\langle k\right|_{s}\right).
\]

As in the previous case, We can compute the probability that the heralded
state is a single photon using $p_{single\, NRD}=\left\langle 1\right|_{s}\hat{\rho}_{heralded\, TD}\left|1\right\rangle _{s}$:

\[
p_{single\, NRD}=\frac{\left(1-\left|\xi\right|^{2}\right)}{p_{trig\, NRD}}\eta_{i}\left(\sum_{n=1}^{\infty}\left|\xi\right|^{2n}n\left(1-\eta_{i}\right)^{n-1}n\eta_{s}^{1}\left(1-\eta_{s}\right)^{n-1}\right)
\]

\[
=\frac{\left(1-\left|\xi\right|^{2}\right)}{p_{trig\, NRD}}\eta_{i}\eta_{s}\left|\xi\right|^{2}\left(\sum_{n=1}^{\infty}n^{2}\left(1-\eta_{i}\right)^{n-1}\left(1-\eta_{s}\right)^{n-1}\left|\xi\right|^{2\left(n-1\right)}\right).
\]

Using $\sum_{n=1}^{\infty}n^{2}x^{n-1}=\sum_{n=1}^{\infty}n\left(n-1\right)x^{n-1}+\sum_{n=1}^{\infty}nx^{n-1}=\frac{2x}{\left(1-x\right)^{3}}+\frac{1}{\left(1-x\right)^{2}}:$ 

\[
p_{single\, NRD}=\left(1-\left(1-\eta_{i}\right)\left|\xi\right|^{2}\right)^{2}\eta_{s}\left(\frac{\left(1+\left(1-\eta_{i}\right)\left(1-\eta_{s}\right)\left|\xi\right|^{2}\right)}{\left(1-\left(1-\eta_{i}\right)\left(1-\eta_{s}\right)\left|\xi\right|^{2}\right)^{3}}\right).
\]

The probability that the heralded state contains multi-photon contamination
is derived as follows. First, we define:

\[
Z_{NRD}=\sum_{m=1}^{\infty}\left\langle 1,m\right|\hat{\rho}\left|1,m\right\rangle 
\]

\[
=\left(1-\left|\xi\right|^{2}\right)\left(\sum_{n=1}^{\infty}\left|\xi\right|^{2n}\sum_{k=1}^{n}C_{n}^{1}\eta_{i}\left(1-\eta_{i}\right)^{n-1}C_{n}^{k}\eta_{s}^{k}\left(1-\eta_{s}\right)^{n-k}\right)
\]

\[
=\left(1-\left|\xi\right|^{2}\right)\left|\xi\right|^{2}\left(\eta_{i}\sum_{n=1}^{\infty}\left|\xi\right|^{2\left(n-1\right)}n\left(1-\eta_{i}\right)^{n-1}\left(1-\left(1-\eta_{s}\right)^{n}\right)\right)
\]

\[
=\left(1-\left|\xi\right|^{2}\right)\left|\xi\right|^{2}\eta_{i}\left[\frac{1}{\left(1-\left|\xi\right|^{2}\left(1-\eta_{i}\right)\right)^{2}}-\frac{\left(1-\eta_{s}\right)}{\left(1-\left|\xi\right|^{2}\left(1-\eta_{i}\right)\left(1-\eta_{s}\right)\right)^{2}}\right]
\]

\[
=\frac{\left(1-\left|\xi\right|^{2}\right)\left|\xi\right|^{2}\eta_{i}\eta_{s}}{\left(1-\left|\xi\right|^{2}\left(1-\eta_{i}\right)\right)^{2}}\left(\frac{1-\left(1-\eta_{s}\right)\left(\left|\xi\right|^{2}\left(1-\eta_{i}\right)\right)^{2}}{\left(1-\left|\xi\right|^{2}\left(1-\eta_{i}\right)\left(1-\eta_{s}\right)\right)^{2}}\right).
\]

Then it follows that the probability that the heralded state contains
multi-photon contamination is given by $p_{multi\, NRD}=\sum_{m=2}^{\infty}\left\langle m\right|_{s}\hat{\rho}_{heralded\, NRD}\left|m\right\rangle _{s}$:

\begin{equation}
p_{multi\, NRD}=\frac{Z_{NRD}}{p_{trig\, NRD}}-p_{single\, NRD}.\label{eq:p_multi_TD-2}
\end{equation}

\[
p_{multi\, NRD}=\eta_{s}\frac{1-\left(1-\eta_{s}\right)\left(\left|\xi\right|^{2}\left(1-\eta_{i}\right)\right)^{2}}{\left(1-\left|\xi\right|^{2}\left(1-\eta_{i}\right)\left(1-\eta_{s}\right)\right)^{2}}-p_{single\, NRD}.
\]

\[
\]

\section*{Appendix B: Chain source derivation}

The exact probability of single-photon emission $q_{chain}$ can be
calculated as follows. The state from a chain of N sources is:

\[
\hat{\rho}_{D}=\sum_{j=0}^{N-1}p_{trig\, D}\left(1-p_{trig\, D}\right)^{j}\hat{\rho}_{j\, D},
\]

where $\hat{\rho}_{j\, D}$ is the state heralded by an individual
HSPS for which the signal arm goes through $j+1$ switches each having transmission $\eta$, and $D$ is the type of detector (TD for threshold
detector or NRD for number resolving detector).

For a number resolving detector:

\[
\hat{\rho}_{j\, NRD}=\frac{\left(1-\left|\xi\right|^{2}\right)}{p_{trig\, NRD}}\eta_{i}\left(\sum_{n=1}^{\infty}\left|\xi\right|^{2n}n\left(1-\eta_{i}\right)^{n-1}\sum_{k=0}^{n}C_{n}^{k}\left(\eta_{s}^{k}\eta^{j+1}\right)\left(1-\left(\eta_{s}\eta^{j+1}\right)\right)^{n-k}\left|k\right\rangle _{s}\left\langle k\right|_{s}\right).
\]

For a threshold detector:

\[
\hat{\rho}_{j\, TD}=\frac{\left(1-\left|\xi\right|^{2}\right)}{p_{trig\, TD}}\left(\sum_{n=1}^{\infty}\left|\xi\right|^{2n}\sum_{p=1}^{n}\sum_{k=0}^{n}C_{n}^{p}\eta_{i}^{p}\left(1-\eta_{i}\right)^{n-p}C_{n}^{k}\left(\eta_{s}\eta^{\mathbf{j}+1}\right)^{k}\left(1-\left(\eta_{s}\eta^{\mathbf{j}+1}\right)\right)^{n-k}\left|k\right\rangle _{s}\left\langle k\right|_{s}\right).
\]

The probability that the MUX source triggers is given by: 

\begin{align*}
p_{trig}^{chain} & =1-\left(1-p_{trig}\right)^{N}.
\end{align*}

The probability of single-photon emission conditioned on the MUX source
triggering is given by: 

\begin{align*}
p_{single\, D}^{chain} & =\frac{1}{p_{trig}^{chain}}\left\langle 1\right|_{s}\hat{\rho}_{D}\left|1\right\rangle _{s}\\
 & =\frac{1}{p_{trig}^{chain}}\sum_{j=0}^{N-1}p_{trig\, D}\left(1-p_{trig\, D}\right)^{j}p_{single\, j\, D},
\end{align*}

where $p_{single\, j\, D}$ is the probability of single-photon emission
for an individual HSPS heralding the state $\hat{\rho}_{j\, D}$ (calculated
in the same way as in Section \ref{sec:The-heralded-single}
but using $\hat{\rho}_{j\, D}$ ). Then we have:

\[
q_{chain\, D}=p_{single\, D}^{chain}\times p_{trig\, D}^{chain}.
\]

The probability of multi photon contamination conditioned on the MUX
source triggering is given by:

\[
\]
\begin{equation}
p_{multi}^{chain}=\frac{1}{p_{trig}^{chain}}\sum_{j=0}^{N-1}p_{trig\, D}\left(1-p_{trig\, D}\right)^{j}p_{multi\, j\, D},\label{eq:pmulti}
\end{equation}

with

\[
p_{multi\, j\, D}=\sum_{m=2}^{\infty}\left\langle m\right|_{s}\hat{\rho}_{j\, D}\left|m\right\rangle _{s}
\]

\[
=Z_{j\, D}-p_{single_{j}\, D},
\]

where $Z_{j\, D}$ is the normalization constant for an individual
HSPS heralding the state $\hat{\rho}_{j\, D}$ (calculated in the
same way as in Sec. \ref{sec:The-heralded-single} but using
$\hat{\rho}_{j\, D}$) Then we can rewrite \ref{eq:pmulti}  as:

\[
p_{multi}^{chain}=\frac{Z_{D}^{chain}}{p_{trig}^{chain}}-p_{single\, D}^{chain},
\]

where

\[
Z_{D}^{chain}=\sum_{j=0}^{N-1}p_{trig\, D}\left(1-p_{trig\, D}\right)^{j}Z_{j\, D}.
\]

\bibliographystyle{unsrt}
\bibliography{multiplexed}

\end{document}